# Towards a Formal Specification Framework for Manufacturing Execution Systems

Maria Witsch and Birgit Vogel-Heuser, *Member, IEEE*

***Abstract*—Manufacturing Execution Systems (MES) optimize production and business processes at the same time. However, the engineering and specification of MES is a challenging, interdisciplinary process. Especially IT and production experts with different views and background have to cooperate. For successful and efficient MES software projects, misunderstandings in the specification process have to be avoided. Therefore, textual specifications need to be complemented by unambiguous graphical models, reducing the complexity by integrating interdisciplinary views and domain specific terms based on different background knowledge. Today's modeling notations focus on the detailed modeling of a certain domain specific problem area. They do not support interdisciplinary discussion adequately. To bridge this gap a novel MES Modeling Language (MES-ML) integrating all necessary views important for MES and pointing out their interdependencies has been developed. Due to its formal basis, comparable and consistent MES-models can be created for specification, standardization, testing, and documentation of MES software. In this paper, the authors present the formal basis of the modeling language and its core notation. The application of MES-ML is demonstrated taking a yogurt production as an example. Finally, the authors give some evaluation results that underline the effectiveness and efficiency of this new modeling approach with reference to four applications in industrial MES-projects in the domain of discrete and hybrid manufacturing.**

## I. Introduction

FOR A competitive production process, vertical integration of the automation layer and the Enterprise Resource Planning (ERP) layer is necessary. Manufacturing Execution Systems (MES) are process-oriented software systems which represent the interface between the automation layer and ERP systems [1], [2]. Linking production lines with business level leads to cost-effective and more efficient production [3]. Automated data transfer between automation and ERP by MES improve production and business processes by avoiding data errors during manual input, accelerating reporting, enabling flexible production planning, and replacing repeated manual operations for example. The profitability of manufacturers is more and more defined by efficient, integrated process optimization from ERP to the field than by highly optimized standalone automation system [4]. The growing importance of knowledge in the production process [5] strengthens the relevance of MES as an enterprise information integration system [1]. To fully exploit the advantages of MES, in the first phases of a MES project, the available or required data have to be identified and the functionality to be provided by the new MES in differentiation to existing IT-systems has to be specified accurately. These specifications have to be easily comprehendible by all participating stakeholders within the interdisciplinary implementation process. As shown in [6], [7], and Section II-B of this paper, no graphical modeling notation supports this specification process adequately at the moment. Therefore, currently in practice, specifications are mostly textual without graphical models. The documents are ambiguous and often not intuitively understandable by people from other disciplines than the author's. For that reason, these documents are not suited for interdisciplinary discussion. For MES customers, this entails missing transparency regarding the concrete functional requirements and selection of MES functions. The required transparency for the integration of a new MES into an existing IT-infrastructure can hardly be achieved without an appropriate graphical model [8]. "The significance of process modeling to industrial informatics is obvious" [9]. Business process modeling supports enterprises in standardization, optimization, and reengineering of their business processes [10]. Modeling of manufacturing processes enables manufacturers to understand and optimize their processes [11] and thereby helps to identify requirements to the MES. Such a graphical modeling language should support the requirement engineering process of MES and document all functional requirements. This functional description should be usable as requirements documentation, test specification and as blueprint for the MES.

The MES modeling language should be based on a formal syntax so that consistency of the MES model can be checked automatically. This increases the model's quality and reliability. A modeling language with formal syntax and semantics is necessary to avoid ambiguities in requirement specification documents or contracts, respectively. A formal syntax constitutes the first step towards a formal semantics definition (such as structural operational semantics), which would enable to add simulation techniques or even formal methods like model checking to a MES modeling tool chain [12]. Moreover, a precise syntax is required to obtain comparable models and compatible tool-implementations.

The MES Modeling Language **(MES-ML)** presented in this paper, was developed to fulfill these requirements and constitutes a very promising approach as industrial applications show (see Section VII). It is integrated in a modeling method to assure model quality and supported by a software prototype to facilitate application (which are both not in the focus of this paper).

Manuscript received September 11, 2011; revised November 16, 2011; accepted January 03, 2012. Date of publication February 03, 2012; date of current version April 11, 2012. This work was supported by the German Ministry of Education and Research (BMBF) under Grant number 01IS09026C. Paper no. TII-11-517.

The authors are with the Institute of Automation and Information Systems, Technische Universität München, 85748 Garching bei München, Germany (e-mail: m.witsch@tum.de; vogel-heuser@ais.mw.tum.de).

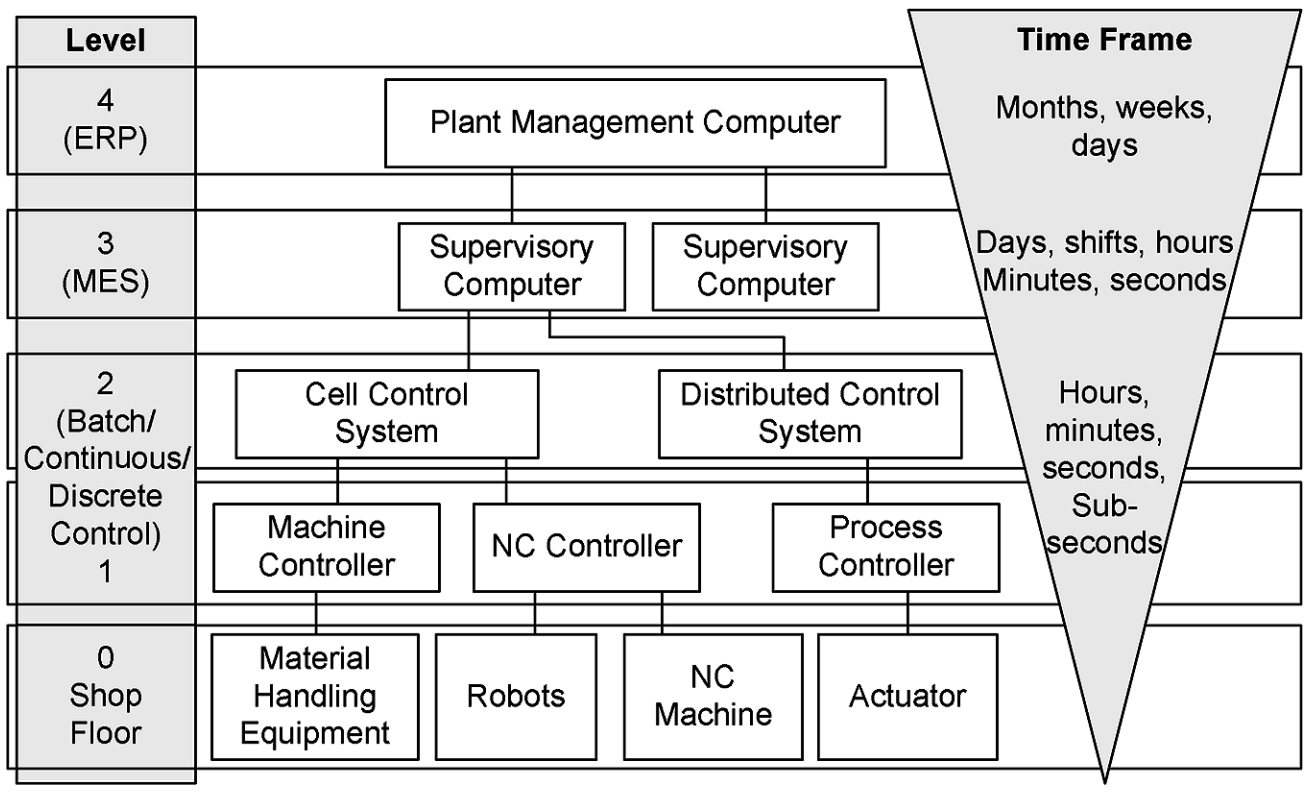

Fig. 1. MES as interface between ERP and control level [15]–[17].

After giving a short overview about MES, characteristics of MES engineering projects and resulting requirements *(Req. 1–Req. 10)*, existing modeling notations are compared to these requirements. A modeling framework closing the identified gaps is presented in Section III and its formalization is given in Section IV. Section V describes the notation for the elements used in the example of Section VI. The evaluation results of four industrial applications are presented in Section VII. Section VIII concludes this paper.

## II. MES, Challenges in MES Engineering and Existing Methods

### A. Manufacturing Execution Systems

In the last few years, different definitions of MES have been developed by industry and academia, e.g., by the Manufacturing Enterprise Solutions Association (MESA) [13] and the Instrumentations, Systems, and Automation Society (ISA) [14]. The wide range of MES functionality defined in [15] can be categorized in production operation management, quality operation management, inventory operation management, and maintenance operation management. While automation usually operates in seconds or milliseconds, ERP acts in middle or long-term range [15].

### B. Challenges in MES Engineering

Fig. 1 shows that the MES connect control systems with ERP. Due to this interface character MES engineering is an interdisciplinary challenge and not realizable by only one domain. Plant engineers, MES engineers, plant manager, production manager, IT engineers, solution developer, etc., need to cooperate to specify the entire requirements of the MES. They provide different details about the processes and have diverse requirements concerning the functionality of the MES. Furthermore, they use different terms and models for describing them. Consequently, it is difficult to get a fruitful discussion in interdisciplinary workshops and to document the consolidated requirements in a roundly accepted and well understandable form. For MES customers, this leads to a suboptimal cooperation and islands of knowledge among responsible co-workers. *(Req. 1: Interdisciplinarily understandable, not complex, and quickly learnable for application in interdisciplinary workshops.)*

Processes, supposed to be monitored by the MES, range from incoming order handling for goods to distribution of finished products [18], [19]. Therefore, business processes, automated production processes, maintenance, and quality processes have to be considered *(Req. 2: Distinction between different kinds of processes and their interaction and time relations.)* These processes are very diverse, complex and can be displayed only in dynamic models. Some of them might be automated and could by integrated in the MES (i.e., automated data exchange). *In this case, an interface definition has to be part of the MES specification document (Req. 3)*. Some of these processes might be executed completely manually. In this case, the workflow and user interactions with the MES have to be defined *(Req. 4: Differentiation between workflow, MES and user activities and illustration of their interaction)*. Hence, the first issue to consider in a MES project is a detailed process analysis [18].

The next step in MES projects is to define the target process and the role of the MES in this process and thereby its required functionality. In most cases MES are integrated in existing IT-systems. Therefore, analysis of the current offered functionality by running systems and a (re)assignment of functional requirements to existing IT-systems and the new MES is essential, especially to avoid redundant functionalities and data as well as unnecessary effort for adaptation of legacy systems *(Req. 5: assignment of functional requirements to existing IT-systems and the new MES)*. Due to the complexity of MES software, models of MES require complexity reducing concepts like a composition/decomposition *(Req. 6: functional definition of the MES with complexity reducing concepts)*. *A direct comparison of current and target process in one diagram is necessary* to make the benefits of the MES implementation visible and identify the indicators for ROI calculation *(Req. 7)*.

The technical system provides the data, which can be used for MES functions. Necessary data points needed for the new MES system have to be identified *(Req. 8: identification of data dependencies)*. Relevant data which are already recorded by the Process Control System (PCS) or provided by the machines themselves and which are integrated into the MES database by import have to be determined, specified, and interpreted. This interpretation is often difficult, as data points are often neither labeled nor self-explanatory. Sometimes automation engineers hide relevant device information to reduce complexity. These data have to be recovered by the control and MES engineer [20]. The basis for the definition and interpretation of data is a precise understanding of the existing technical system or the part to be optimized and the existing IT infrastructure *(Req. 9: illustration of the technical system and IT infrastructure)*. By analyzing the existing IT infrastructure potentially free capacity usable for the new system can be identified and additionally required hardware can be defined *(Req. 10: illustration of the relationship between MES functionality and IT-infrastructure)* [8].

### C. Existing Methods

Some MES vendors have developed their own proprietary modeling notations where each element represents one module of their specific MES software solution. These notations are only applicable after provider decision and limited to the functional view of their MES. Others use modeling notations from

TABLE I
COMPARISON OF DIFFERENT MODELING NOTATIONS

| | REQUIREMENT | BPMN 2.0 [32,40] | Flowcharts [24] | Petrinets [21,22,26,27] | UML 2.x [23,28] | SysML [29- 31] |
|---|---|---|---|---|---|---|
| 1 | **Interdisciplinarily understandable, not complex and quickly learnable for application in interdisciplinary workshops** | (✓)* | ✓ | × | × | × |
| 2 | (a) Distinction between different kinds of processes | ✓ | ✓ | ✓ | ✓ | ✓ |
| | (b) and their interaction and time relations | ✓ | × | × | ✓ | ✓ |
| 3 | interface definition | × | × | × | ✓ | ✓ |
| 4 | (a) Differentiation between workflow, MES and user activities | × | × | × | × | × |
| | (b) illustration of their interaction | × | × | × | × | × |
| 5 | assignment of functional requirements to existing IT-systems and the new MES | ✓ | × | × | ✓ | ✓ |
| 6 | functional definition of the MES with complexity reducing concepts | ✓ | ✓ | ✓ | ✓ | ✓ |
| 7 | comparison of current and target process in one diagram | × | × | × | × | × |
| 8 | identification of data dependencies | ✓ | ✓ | × | ✓ | ✓ |
| 9 | illustration of the technical system and IT-infrastructure | × | × | × | × | ✓ |
| 10 | illustration of the relationship between MES functionality and IT-infrastructure | × | × | × | × | × |

** The core set of BPMN elements can be judged as not complex and easily learnable in contrast to the extended set.*

associated domains to discuss parts of the requirements concerning the integration of MES into the business process or to describe the behavior of flexible manufacturing systems using Petri nets [21], [22]. Models like UML Activity Diagrams [23] or Flowcharts [24] are a common notation for modeling the production process (e.g., [25]).

All these modeling notations are usually limited to one view of the plant or production or business process level and do not consider automation as well as technical requirements. In addition, they are not intuitively understandable, mostly complex, and therefore not suited for interdisciplinary discussion which is the most important requirement for a MES modeling language [6], [7].

Table I shows further gaps by comparing the requirements derived in Section II-A and the named modeling notations:

Research shows that existing modeling notations already fulfill different requirements. Therefore, one of the goals when developing a MES specific modeling language was to use a state-of-the-art modeling language. This allows modeling with known notation and at least partly defined syntax and semantics.

The Business Process Model and Notation (BPMN) [32], developed by the Object Management Group (OMG), appears to be the most powerful modeling notation for business process modeling which is interdisciplinary understandable [33]. It gains importance and is supported by an increasing number of software tools. The previous version [34] was already considered as the *de-facto* standard for business process modeling [35]. Over the last years, numerous applications and adaptations of the BPMN focusing, e.g., on service management processes [36], rich Internet applications [37], bank loan [38] and healthcare processes [39] to vehicle development [40] and manufacturing processes [41] are reported. Especially the application for business as well as manufacturing processes indicates applicability for MES engineering processes. Even though the BPMN's scope is widespread, it is an intuitively understandable language and was therefore chosen for modeling of production processes, MES- and IT-functionality.

To apply the BPMN for modeling the production process and MES and through this realize an interdisciplinary MES specification process, elements of the BPMN have to be selected [42] and their semantics in the context of MES have to be defined.

The modeling concept presented in the following closes the gaps between the identified requirements and existing languages by integrating the necessary views for interdisciplinary MES engineering. Besides MES specific requirements presented before, the MES-ML considers usability aspects for modeling notations given in [43] and [44].

## III. MES-Modeling Language

According to the requirements the MES-ML provides three views, each addressing people with different expertise about the modeled system. The first view describes the *technical system* with its different components and its IT-infrastructure in a simple but comprehensive way and in hierarchical order. To collect and use this information about the technical system in interdisciplinary discussions, a simple static model (hierarchical tree structure) of the technical system is sufficient. The **Technical System Model** ($M_{\mathrm{TS}}$) represents the technological structure of the plant as a whole, down to atomic function units at the most detailed level. For the definition of interfaces between the MES and the technical system, more details about existing data, data quality and semantics of data are helpful. Therefore, each functional unit can be detailed by attributes concerning its quality, metadata, and semantics.

The second view specifies the *production process*. The **Production Process Model** ($M_{\mathrm{PP}}$) represents the workflow of the production and business processes and additional processes such as maintenance or quality management process steps modeled with a BPMN variant.

The *MES/IT functional view*, describing properties and functionalities the MES should implement, is supported by the **MES/IT-Model** ($M_{\mathrm{MES}}$). It allows describing functions and processes realized by the MES and displays the functions of MES embedded in interacting IT systems. This is also modeled using an adapted BPMN.

Table II maps the requirements given in Table I to the selected and adapted modeling elements and developed MES-ML concepts.

Each of these three models realize a specific view on the MES and can be created independently.

The integration of these three views and transparent identification of their interdependencies (Req. 2, 3, 4, 8, 10) is realized by the **MES-ML Linking Model** $(M_{\mathrm{LK}})$. It defines three types of links: **Data Transfer** (Req. 3, 8), **Equivalence** (Req. 2, 4), and **Deployment** (Req. 10).

TABLE II
MAPPING OF LANGUAGE REQUIREMENTS TO MES-ML CONCEPTS/ELEMENTS

| REQ. NO. | MES-ML CONCEPTS/ELEMENTS |
|---|---|
| 2 | Integration of activities in the sequence flow, definition of global activities, link events, intermediate events, gateways |
| 3 | Connector type as attribute |
| 4 | (a) Separation of production process and MES model and colored distinction in combination with the execution type<br>(b) Sequence, data and message flows and reference elements |
| 5 | Pools and lanes |
| 6 | Start-/stop-events, sequence flow, sub processes, link events |
| 7 | Attribute requirement status |
| 8 | Data objects, data stores, data flows, message flows |

### *A. Data Transfer Link*

A Data transfer models that there is or should be an interface (implemented or to be implemented by software) between two elements of different models $(M_{\mathrm{MES}}, M_{\mathrm{PP}}, M_{\mathrm{TS}})$. By this a dependency between two functions/actions is modeled. This kind of interaction is semantically equivalent to message flows inside the MES and the production process model but without a graphical representation. This link shows the MES's impact on the production process and vice versa. These links are 1:1 relations.

### *B. Equivalence Link*

During modeling, the MES situations can occur where a process step of the production and business process has to be modeled without knowing if it is solely a production step, if it interacts with the MES on a more detailed level or if it can be divided into steps executed by the IT and the plant. At least, if the MES is not implemented for an existing plant but is projected in the early engineering phases of a plant on a less detailed level regarding the description of the MES and production process, there are process steps, whose executing system (automation system or MES) is not specified at this time. These process steps are of interest in both the MES functional model and the production process model. In this case, it is possible to model that the production process-activity in the MES model and the MES-activity in the production process model represent the same process step through the activity reference element and an equivalence link.

### *C. Deployment Link*

The third link type guides the modeler and any user of the specification through the different models and helps to understand the plant's entire extent, its production process and the part of the MES in between. It displays which production process takes place on which part of the plant and which MES, module or function has to be realized by which IT-infrastructure component.

These extensions of the BPMN leads to several benefits in comparison to a single use of one of the compared modeling languages in Table I. The differentiation by color (gray and white background) is necessary to distinct between IT and production processes. Preceding works showed that an integration of both views in one model leads to very complex and huge models on paper walls. If the production workflow was taken as leading process and single MES functions were just integrated, it was not possible to completely define the structure and processes of the MES. Therefore, it was impossible to get a complete understanding of how a process works. If the MES functional structure and on more detailed level its business logic was taken as leading process and just single production activities were integrated a good understanding of how the production works could not be achieved. The content in which this activity takes place, what happens before and afterwards is necessary to define an activity unambiguously. Nevertheless, it stays inevitable to model the interactions and interdependencies between the two processes. Hence, a link-concept between the models was integrated in the MES-ML. Links provide the necessary information about the interdependencies without interrupting or complicating the sequence flow inside the model. As processes of both disciplines can be modeled adequately, their motivation to participate in the requirements engineering process is higher. Additionally, the Technical System Model is necessary to provide the information about available sensors and actuators and facilitate especially the communication with electrical engineers to define the interfaces to the machines.

The abstract syntax of the MES Specification Model $(\mathrm{MES}_{\mathrm{Spec}})$, which aggregates these three views and links will be introduced in the following in a formal way. To distinguish between element connections within one model and connections across different models, the first will be named "connection," the second "link."

## IV. FORMALIZATION OF THE MES-MODELING LANGUAGE

A **MES Specification Model** is defined as a tuple in the below equation[1]

$$\mathrm{MES}_{\mathrm{Spec}} = \langle M_{\mathrm{MES}}, M_{\mathrm{PP}}, M_{\mathrm{TS}}, M_{\mathrm{LK}} \rangle \tag{1}$$

containing one MES functional model $|M_{\mathrm{MES}}| = 1$, one production process model $|M_{\mathrm{PP}}| = 1$, one technical system model $|M_{\mathrm{TS}}| = 1$ and a MES-ML linking model $|M_{\mathrm{LK}}| = 1$.

A **Technical System Model** is a tuple

$$M_{\mathrm{TS}} = \langle p, \mathrm{AR}, U, SN, \mathrm{UDL} \rangle \tag{2}$$

with $p$ as a plant, AR as a non-empty set of plant areas $\mathrm{AR} = \{\mathrm{ar}_1, \ldots, \mathrm{ar}_n\}, U$ as a non-empty set of units $U = \{u_1, \ldots, u_n\}$ and SN as a non-empty set of signals $\mathrm{SN} = \{\mathrm{sn}_1, \ldots, \mathrm{sn}_n\}$ as well as UDL as a set of user-defined layers $\mathrm{UDL} = \{\mathrm{udl}_1, \ldots, \mathrm{udl}_n\}$. These sets belong to the plant which represents the root-node. Signals are attributes of units and therefore can only exist, if units are modeled.

[1]Most important equations are formatted as equations and numbered in the following.

A **Production Process Model** is defined as a tuple

$$M_{\text{PP}} = \langle \text{FO}_{\text{PP}}, \text{DO}_{\text{PP}}, \text{CO}_{\text{PP}}, \text{AF}_{\text{PP}} \rangle \quad (3)$$

and consists of flow objects $|\text{FO}| \geq 3$, data objects $|\text{DO}| \geq 0$, connecting objects $|\text{CO}| \geq 2$, and artifacts $|\text{AF}| \geq 0$. In the following definitions each occurrence of the variable $\boldsymbol{n}$ represents a different, independent variable $n \in \mathbb{N}^+$.[2]

**Flow objects** $\text{FO} = \langle A, E, G, \text{RE}_A, \text{RE}_{\text{SN}} \rangle$ consist of a non-empty set of activities $A = \{a_1, \ldots, a_n\}$ with a set of events $E = \{e_1, \ldots, e_n\}, |E| \geq 2$, a set of gateways $G = \{g_1, \ldots, g_n\}$, a set of Activity Reference Elements $\text{RE}_A = \{\text{re}_{A,1}, \ldots, \text{re}_{A,n}\}$, and a set of Signal Reference Elements $\text{RE}_{\text{SN}} = \{\text{re}_{\text{SN},1}, \ldots, \text{re}_{\text{SN},n}\}$.

**Data objects** are defined as a tuple $\text{DO} = \langle \text{SDO, MDO, DS} \rangle$ with $\text{SDO} = \{\text{sdo}_1, \ldots, \text{sdo}_n\}$ as a set of single data objects, $\text{MDO} = \{\text{mdo}_1, \ldots, \text{mdo}_n\}$ as a set of multi data objects and $\text{DS} = \{\text{ds}_1, \ldots, \text{ds}_n\}$ as a set of data stores.

The tuple $\text{CO} = \langle \text{SF}, \text{IF}, \text{AS} \rangle$ defines **connecting objects** which consist of a set of sequence flows $\text{SF} = \{\text{sf}_1, \ldots, \text{sf}_n\}$, a tuple information flows $\text{IF} = \langle \text{MF}, \text{DF} \rangle$ (containing a set of message flows $\text{MF} = \{\text{mf}_1, \ldots, \text{mf}_n\}$ and data flows $\text{DF} = \{\text{df}_1, \ldots, \text{df}_n\}$), and a set of associations $\text{AS} = \{\text{as}_1, \ldots, \text{as}_n\}$. **Artifacts** are defined as a tuple $\text{AF} = \langle \text{TA}, \text{GR} \rangle$ that consists of text annotations $\text{TA} = \{\text{ta}_1, \ldots, \text{ta}_n\}$, and groups $\text{GR} = \{\text{gr}_1, \ldots, \text{gr}_n\}$.

For the definition of links between the models, elements of flow objects and reference elements are most important and will therefore be presented in more detail below.

For MES specification, a distinction between different kinds of activities is helpful. The production process and the MES interact permanently. Therefore, an activity is defined as a tuple

$$a_i = \langle \text{sp}, q_a, \text{rt}_a, \text{rst}_a, n_{a,\text{isf}}, n_{a,\text{osf}}, n_{a,\text{iif}}, n_{a,\text{oif}} \rangle \quad (4)$$

with one optional set of sub processes $\text{sp} = \langle \text{FO}_{\text{PP}}, \text{DO}_{\text{PP}}, \text{CO}_{\text{PP}}, \text{AF}_{\text{PP}} \rangle$ contained in the activity, $\text{q}_\text{a}$ as attribute execution type with $\text{q}_\text{a} \in \text{Q}_\text{a} = \{\text{Manual}, \text{Automatic}\}$, $\text{rt}_\text{a}$ as attribute repetition type with $\text{rt}_\text{a} \in \text{RT}_\text{a} = \{\text{none}, \text{sequential}, \text{parallel}\}$, $\text{rst}_\text{a}$ as attribute requirement status with $\text{rst}_\text{a} \in \text{RST}_\text{a} = \{\text{ToImplement}, \text{Implemented}, \text{Excluded}\}$, and $\text{n}_{\text{a,isf}}$ as number of incoming sequence flows, $\text{n}_{\text{a,osf}}$ as number of outgoing sequence flows, $\text{n}_{\text{a,iif}}$ as number of incoming information flows, and at least $\text{n}_{\text{a,oif}}$ as number of outgoing information flows. Manual activities in the production process model represents activities, performed by workers during the manufacturing or production process including maintenance or quality proving tasks. Automatic activities symbolize tasks automatically executed by the control software on the machines.

Production process activities described in the production process model affect activities in the MES functional model and *vice versa*. Actions logically belonging to the MES but having a huge impact on the production process, can be modeled as activity reference element in the production process model.

Activity reference elements are defined as function $\lambda_a$

$$\begin{aligned} \text{RE}_A = \langle \lambda_a \rangle, \lambda_a : A \rightarrow A_{\text{ref}} \text{ with } A_{\text{ref}} \subseteq A \\ \forall a_x \in A_{\text{ref}}, \exists \text{re}_{a,x} \in \text{RE}_A : \lambda_{a,\text{re}_x} \rightarrow a_x. \end{aligned} \quad (5)$$

A signal $\text{sn}_x$ defined in the Technical System Model representing a data source for a production process activity $\text{a}_{\text{pp,x}}$ can be displayed in the Production Process Model by a Signal Reference Element $\text{re}_x$, defined as a function $\lambda_{\text{sn},\text{re}_x}$

$$\begin{aligned} \text{RE}_{\text{SN}} = \langle \lambda_{\text{sn}} \rangle, \lambda_a : \text{SN} \rightarrow \text{SN}_{\text{ref}} \text{ with } \text{SN}_{\text{ref}} \subseteq \text{SN} \\ \forall \text{sn}_x \in \text{SN}_{\text{ref}}, \exists \text{re}_{\text{sn},x} \in \text{RE}_{\text{SN}} : \lambda_{\text{sn},\text{re}_x} \rightarrow a_x. \end{aligned} \quad (6)$$

An event is defined as a tuple

$$e_i = \langle q_e, b_e \rangle \quad (7)$$

with $q_e$ as an attribute event execution type and $\text{q}_\text{e} \in \text{Q}_\text{e} = \left\{ \begin{matrix} \text{Start, Stop, IntermediateInterrupting(II),} \\ \text{IntermediateNonInterrupting(INI)} \end{matrix} \right\}$, and $b_e$ as an attribute behavior type and $b_e \in B_e = \{\text{Timer}, \text{Error}, \text{Link}\}$.

A gateway is defined as a tuple

$$g_i = \langle q_g, b_g, n_{g,\text{isf}}, n_{g,\text{osf}}, n_{g,\text{iif}}, n_{g,\text{oif}} \rangle. \quad (8)$$

$q_g \in Q_g = \{\text{Exclusiv}, \text{Inclusiv}, \text{Parallel}\}$ represents the execution type of gateways, $b_g$ describes the behavior type with $b_g \in B_g = \{\text{Split}, \text{Merge}\} n_{g,\text{isf/osf/iif/oif}}$, gives the number of incoming and outgoing sequence and message flows as defined for activities. According to the parameter value of each defined type of gateway there are additional requirements:

1) if $b_g = \text{Split}$ and $q_g = \text{Exclusiv}$ or $\text{Parallel} \rightarrow n_{g,\text{isf}} = 1$ and $n_{g,\text{osf}} \geq 2$;
2) if $b_g = \text{Split}$ and $q_g = \text{Inclusiv} \rightarrow n_{g,\text{isf}} = 1$ and $n_{g,\text{osf}} \geq 3$;
3) if $b_g = \text{Merge}$ and $q_g = \text{Exclusiv, Inclusiv}$ or $\text{Parallel} \rightarrow n_{g,\text{isf}} \geq 2$ and $n_{g,\text{osf}} = 1$.

In addition $n_{g,\text{idf}} = 0$ and $n_{g,\text{odf}} = 0$ is true for all gateways.

To distinguish between elements from $\text{M}_{\text{MES}}$ and $\text{M}_{\text{PP}}$ we use the indices MES and PP for each element in the following.

A **MES functional model** is a tuple

$$M_{\text{MES}} = \langle \text{FO}_{\text{MES}}, \text{DO}_{\text{MES}}, \text{CO}_{\text{MES}}, \text{AF}_{\text{MES}}, \text{SL}_{\text{MES}} \rangle. \quad (9)$$

For $\text{FO}_{\text{MES}}, \text{DO}_{\text{MES}}, \text{CO}_{\text{MES}}, \text{AF}_{\text{MES}}$ and their entire sub elements the same definitions as shown in the last section for $\text{FO}_{\text{PP}}, \text{DO}_{\text{PP}}, \text{CO}_{\text{PP}}, \text{AF}_{\text{PP}}$ in the production process model are valid. **Swimlanes** are defined as a tuple $\text{SL} = \langle P, L \rangle$ with pools $P = \{p_1, \ldots, p_n\}, |P| \geq 1$ and lanes $L = \{l_1, \ldots, l_n\}, |L| \geq 1$.

Pools and lanes are elements to structure the diagrams and to define which system is responsible for the execution of the modeled processes and standalone functionalities. The hierarchical structure of computer integrated manufacturing [13] systems can be displayed by vertical alignment of the pools.

An activity in the MES model is defined as a generic term for a software function that shall be performed by the MES or another system depending in which pool the activity is located. $A_{\mathrm{MES}}$ with $q_a = \mathrm{Manual}$ represents user tasks. They are typical "workflow" tasks where a human performs the task with the assistance of the MES software application.

[2]To reduce the complexity and to enhance the readability of the definitions, the authors do not write down the index for each element, hence FO is written instead of $\text{FO}_{\text{PP}}$ and so on also for all following definitions in Section IV.

### A. MES-ML Linking Model

The **MES-ML Linking Model** is defined as a non empty set of links

$$M_{\mathrm{LK}} = \langle \mathrm{lk}_1, \mathrm{lk}_2, \ldots, \mathrm{lk}_\mathrm{n} \rangle. \quad (10)$$

A link connects two objects from different views (models). It is possible to link every object from a view to an object from another view. Links between certain object types are not allowed. For this reason, we subsequently define how well-formed links look like and discuss the semantics of links between the views and special object types. In general, gateways $(G_{\mathrm{PP}}, G_{\mathrm{MES}})$, connecting objects $(\mathrm{CO}_{\mathrm{PP}}, \mathrm{CO}_{\mathrm{MES}})$, link events and text annotations from the MES functional model as well as from the production process model cannot be linked to elements in other models.

$O$ represents the set of elements including every element from the different models that can be linked to elements from other models $O = \{O_{\mathrm{MES}} \cup O_{\mathrm{PP}} \cup O_{\mathrm{TS}}\}$ and the view of an object $o \in O$ is given by the function

$$\mathrm{view}(o) = \begin{cases} \mathrm{MES}, & \text{if } o \in O_{\mathrm{MES}} \\ \mathrm{PP}, & \text{if } o \in O_{\mathrm{PP}} \\ \mathrm{TS}, & \text{if } o \in O_{\mathrm{TS}} \end{cases}.$$

The elements that can be the start or end of a link are defined as tuples,

$$O_{\mathrm{MES}} = \{P_{\mathrm{MES}}, L_{\mathrm{MES}}, A_{\mathrm{MES}}, E_{\mathrm{MES}}, \mathrm{GR}_{\mathrm{MES}}, D_{\mathrm{MES}}\},$$
$$O_{\mathrm{PP}} = \{A_{\mathrm{PP}}, E_{\mathrm{PP}}, \mathrm{GR}_{\mathrm{PP}}, D_{\mathrm{PP}}\} \text{ and}$$
$$O_{\mathrm{TS}} = \{\mathrm{AR}, U, \mathrm{SN}, \mathrm{UDL}\}.$$

It is only possible to link one element to another one. By this we define that for linking a number of elements, they have to be grouped beforehand, and linked to each other by connecting these groups. In addition, only elements from different models can be linked so that for two elements $a, b\{(a,b) \in O \times O |\ \mathrm{view}(a) \neq \mathrm{view}(b)\}$ holds.

A link is defined as a tuple

$$\mathrm{lk}_i = \langle c, \alpha, \beta, t_{\mathrm{lk}} \rangle \quad (11)$$

with $c$ as a connector type that has to be defined by the modeler. In the tool implementation some commonly used interface types such as OPC are predefined. Additionally, the user can define further connection types. $\alpha$ and $\beta$ are functions that define the start and end of a link $\alpha_{\mathrm{lk}} : O \to a$ and $\beta_{\mathrm{lk}} : O \to b$ with $a \neq b$. The link can thereby be defined as directed from $a$ to $b$. The MES-ML link model defines three types of links $t_{\mathrm{lk}} \in \{\mathrm{Datatransfer}, \mathrm{Equivalence}, \mathrm{Deployment}\}$. A combination of the link type and the direction of the link leads to restrictions of the linkable objects:

$\forall \mathrm{lk} \in M_{\mathrm{LK}}$ with $\alpha(\mathrm{lk}) = a, \beta(\mathrm{lk}) = b$ the following conditions have to be satisfied:

- $t_{\mathrm{lk}} = \mathrm{DataTransfer} \Rightarrow \mathrm{view}(a) = \{A_{\mathrm{MES}} \cup E_{\mathrm{MES}} \cup \mathrm{DO}_{\mathrm{MES}} \cup \mathrm{SL}_{\mathrm{MES}} \cup A_{\mathrm{PP}} \cup E_{\mathrm{PP}} \cup \mathrm{DO}_{\mathrm{PP}} \cup O_{\mathrm{TS}}\} \wedge \mathrm{view}(b) = \{A_{\mathrm{MES}} \cup E_{\mathrm{MES}} \cup \mathrm{DO}_{\mathrm{MES}} \cup \mathrm{SL}_{\mathrm{MES}} \cup A_{\mathrm{PP}} \cup E_{\mathrm{PP}} \cup \mathrm{DO}_{\mathrm{PP}} \cup O_{\mathrm{TS}}\}$, the data transfer link can be used to connect activities, events, data objects, swimlanes and all technical system objects.
- $t_{\mathrm{lk}} = \mathrm{Equivalence} \Rightarrow \mathrm{view}(a) = O_{\mathrm{MES}} \wedge \mathrm{view}(b) = O_{\mathrm{PP}} \wedge ((a \in A_{\mathrm{MES}} \wedge b \in \mathrm{RE}_{\mathrm{PP},A}) \vee (a \in A_{\mathrm{PP}} \wedge b \in \mathrm{RE}_{\mathrm{MES},A}))$ the equivalence link between MES and PP model can exclusively be defined between activities and activity reference elements.
- $t_{\mathrm{lk}} = \mathrm{Equivalebce} \Rightarrow \mathrm{view}(a) = O_{\mathrm{TS}} \wedge (\mathrm{view}(b) = O_{\mathrm{PP}} \vee \mathrm{view}(b) = O_{\mathrm{MES}}) \wedge (a \in \mathrm{SN} \wedge b \in \mathrm{RE}_{\mathrm{SN}})$, the equivalence link between MES or PP and TS model can only be defined between signals and signal reference elements.
- $t_{\mathrm{lk}} = \mathrm{Deployment} \Rightarrow (\mathrm{view}(a)) = O_{\mathrm{MES}} \vee \mathrm{view}(a) = O_{\mathrm{PP}}) \wedge \mathrm{view}(b) = O_{\mathrm{TS}}$, only objects from the process models (MES and PP) can be deployed to the technical system.
- $t_{\mathrm{lk}} = \mathrm{Deployment} \Rightarrow ((\mathrm{view}(a) = O_{\mathrm{MES}} \vee \mathrm{view}(a) = O_{\mathrm{PP}}) \wedge \mathrm{view}(b) = O_{\mathrm{TS}}) \wedge (a \in \{A_{\mathrm{PP}} \cup A_{\mathrm{MES}}\} \wedge b \in \{\mathrm{AR} \cup U \cup \mathrm{UDL}\})$, activities can be deployed to areas, units and user-defined l*ayers*, signals cannot execute an activity. If two elements are linked by an equivalence relation, the two elements must have the same name.

## V. Notation of MES-Modeling Language Used in the Example

The notation of the technical system model is as simple as the model itself. Each element plant, area, unit and signal has a unique name which represents the element in the hierarchical tree. The elements are connected by dotted lines. To underline the type of each element, icons can be added in front of an element.

The notations of MES- and production process model constitutes of nearly the same elements. In general, all MES-model elements have a white and production process model elements a gray background. Only activity reference elements, referencing to their original element in the respective other model look like their original element. Table III shows the notation for Activities and Reference Elements.

For all other elements only the MES model notation will be presented in Table IV. Artifacts are not used in the example and therefore not presented.

## VI. Specification of a Yogurt Production Plant

Fig. 2 shows parts of a laboratory yogurt production plant model to give an impression of how the MES-ML models look like in practical use and how links are applied.

The model is organized in a hierarchical structure. Beginning with a production process overview (level 0), activities are recursively refined and detailed in different diagrams. Due to the resulting hierarchical structure only detailing of process steps required for MES specification is necessary. Nonetheless, superordinate diagrams contain a complete process.

At the top left side, the top level of the production process model is shown. The yogurt production process can be clustered in five process steps. It is initiated by the start event with the manual sub process activity "Setting up the plant." The following steps "Prepare milk," "Produce Yogurt," and "Bottling" are mostly automated process steps and, therefore, modeled as automatic production process activities. The end event marks the end of the process. The manual sub process activity "Quality Test" is not integrated in the sequence flow. This indicates that its subprocess or subprocesses are function calls triggered by link events or start events triggered by a message or data flow. Such subprocesses can be called from everywhere in the production process model. Each of the subprocess activities is detailed in a separate subprocess diagram. Only the two subprocess diagrams "Prepare Milk" and "Quality test" are displayed here exemplarily.

TABLE III
ACTIVITIES AND REFERENCE ELEMENTS

| Element | Notation for Production Process Model | Notation for MES-Model |
|---|---|---|
| Activities: $a_i \in A$, $q_a = undefined$ | 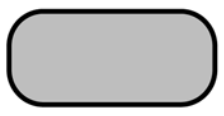 | 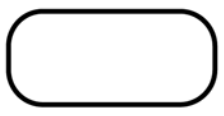 |
| $a_i \in A$, $q_a = manual$ | 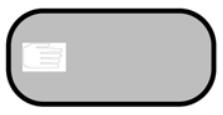 | 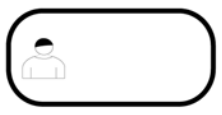 |
| $a_i \in A$, $q_a = automatic$ | 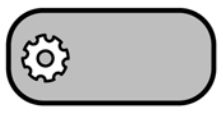 | 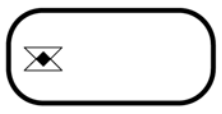 |
| Reference Elements: $re_A \in RE_A$, ($q_a = undefined$) | 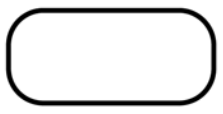 | 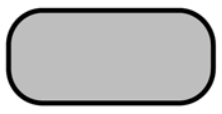 |
| $re_{SN} \in RE_{SN}$ | 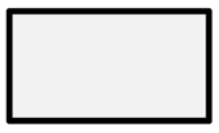 | 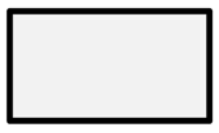 |

TABLE IV
ELEMENTS OF THE MES-MODEL FOR THE YOGURT PRODUCTION EXAMPLE

| Element | Notation |
|---|---|
| Events $e_i \in E$ | 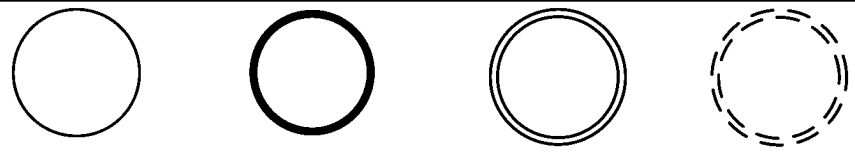 $q_e = Start$; $q_e = Stop$; $q_e = II$; $q_e = INI$ 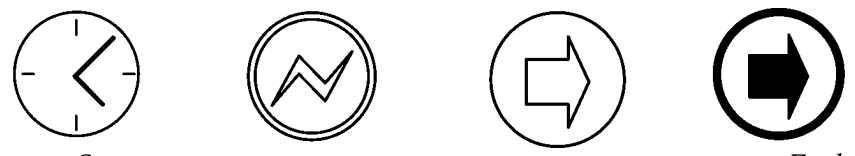 $q_e = Start$, $b_e = Timer$; $q_e = II$, $b_e = Error$; $q_e = Start$, $b_e = Link$; $q_e = End$, $b_e = Link$ |
| Gateways $g_i \in G$ | 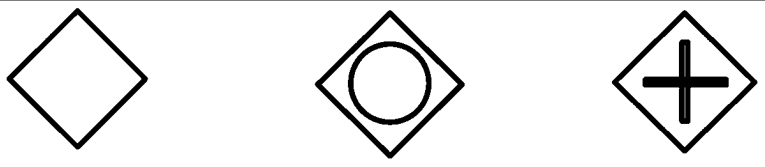 $q_g = Exclusive$; $q_g = Inclusive$; $q_g = Parallel$ |
| Connecting Objects | 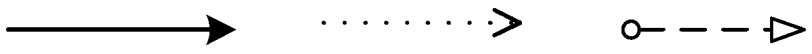 $sf_i \in SF$; $df_i \in DF$; $mf_i \in MF$ |
| Data Objects | 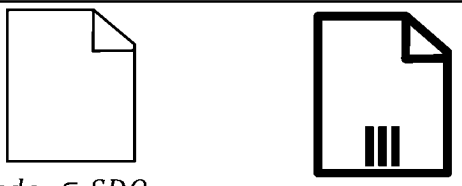 $sdo_i \in SDO$; $mdo_i \in MDO$ |
| Pools and Lanes $p_i \in P$, $l_i \in L$ | 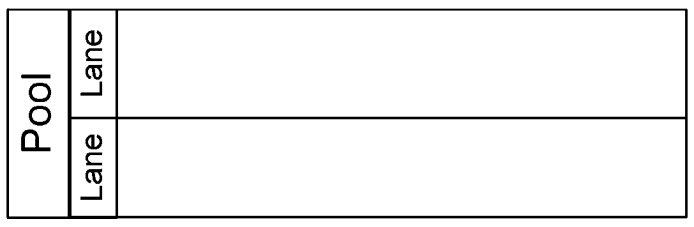  |

On the middle right side, parts of the technical system model are given (hatched area). It contains one workstation, two filling stations, a logistics system and several automation devices. Additionally, the workstation includes different tanks. The attributes of "Tank 101" are displayed with all signals named according to a given process and instrumentation diagram (not part of the MES-ML).

The extract of the MES functional model (dotted area) shows the main MES and IT functions and their superordinated systems: ERP, MES and Process Control System (PCS). As expressed by two separate lanes in the MES pool, the MES is divided into the production management and process data acquisition (PDA) module. Thereby, the overall structure of the MES is defined. The interdependencies between the different IT functions are modeled by message flows. The MES function "Quality Test" is decomposed into two subprocess activities "Collect Test Results" and "Create Sample." Each of those activities is further detailed in a separate diagram and contains further subprocess-activities. Programmers can choose which level to take as base for programming of the MES. This increases the acceptance of the models as requirements specification. Complete models documenting implemented MES which are defined as standard applications can be used for knowledge management and knowledge transfer. Nearly identical elements for IT- and process-description reduce the complexity of the modeling language.

The partial model entitled "Level 2: Quality Test-Create Sample" exemplarily contains two production process activity references "Generate signal sample taking" and "Print Label." While the first is done manually, the latter is executed automatically. Integration of reference elements enables the graphical modeling of MES-production process interaction and interdependencies concerning data and actions.

The black curved lines represent links, which are normally not graphically shown in the model but given as properties in a tool implementation. A detailed explanation of each link is given in the box attached to the lines. Links enable modeling of interdependencies concerning actions, data and events and lead to high transparency regarding interaction between processes and IT. Inside the MES-ML prototype they allow jumping between the views and combination of different views in parallel windows, for example, regarding production process model level 1 "Quality Test" in parallel to MES/IT model level 2 "Collect Test Results" coming from the equivalence link and drop down to one deeper level in the MES/IT model.

## VII. EVALUATION OF THE MES-ML

The MES-ML supported by a software prototype has been evaluated in real specification processes of large enterprises from beverage, automotive and electronic component manufacturing. All evaluations started with a two-day workshop (using the software prototype for direct visualization of the discussed processes and requirements) and were followed by individual further use.

Table V shows the key figures of the evaluation projects.

The applications demonstrate multiple advantages. Complementary to the workshops ten face-to-face expert interviews with the project leading workshop participants were conducted.

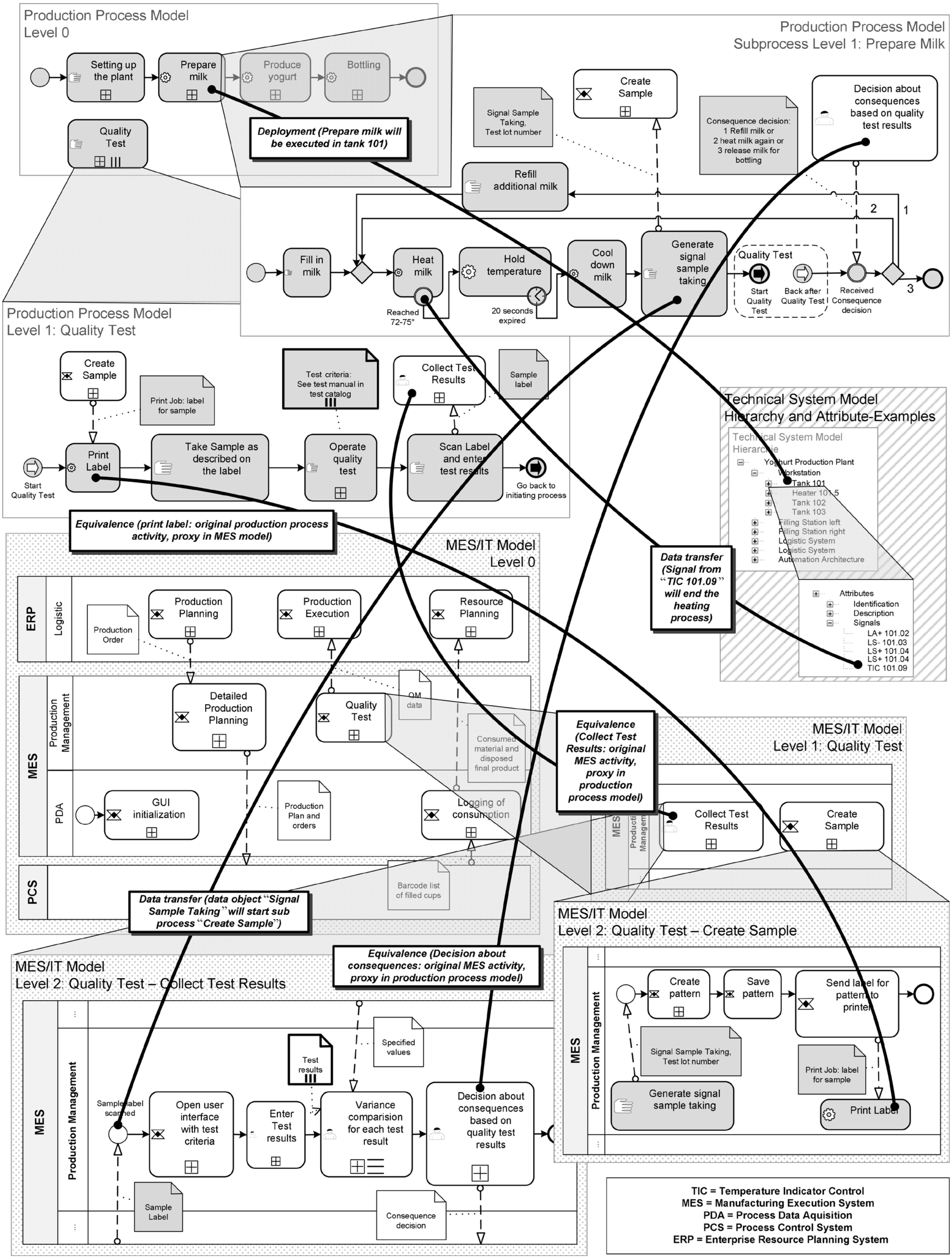

Fig. 2. Extracts of the yogurt production MES-ML-Models with visualization of selected links.

Within these interviews quantitative as well as qualitative data were collected. All participants pointed out that they entirely comprehend the MES-ML models during the workshops independently of their previous knowledge and background. During modeling of production and business processes in workshops they already gain a better understanding of the complex processes and identify process steps to be optimized.

TABLE V
INDUSTRIAL APPLICATIONS OVERVIEW

| **Background of the enterprise** | Automotive | Beverage (soft drinks) | Beverage (dessert) | Electronic Component Manufacturing |
|---|---|---|---|---|
| **Characteristics of evaluation project** | Requirement engineering for MES for replacement of an existing MES and group wide, international standardization of MES | First MES implementation, process analysis and requirement engineering, interaction with existing ERP system | Optimization of IT-systems with MES functionalities | First MES implementation, process analysis and requirement engineering, integration in SAP |
| **Participants** | 4 | 5 | 9 | 12 |
| **Background of participants (quality-management = QM)** | IT | IT, production, management, QM, controlling | IT, production, management, QM, mechanical engineering | IT, production, QM, controlling, mechanical engineering |
| **Modeled diagrams (activities; all other elements) in the 2-days workshop** | **MES**: 18 (>190;>260)<br>**PP**: 1 (8; 11)<br>**TS**: 0 | **MES**: 13 (>80; >120)<br>**PP**: 10 (>70; >80)<br>**TS**: 1 | **MES**: 9 (>60; >80)<br>**PP**: 15 (>80; >100)<br>**TS**: 1 | **MES**: 15 (>80; >120)<br>**PP**: 1 (14; 20)<br>**TS**: 1 |
| **Further use by enterprise** | Models used within specification documents and discussion with MES supplier, Test specification, standardization of MES functionalities, **MES-ML as standard for all following MES projects** | Business and production process redesign (due to identified space for improvements by using MES-ML), reassignment of IT functionalities between ERP and MES (due to identified errors in the implementation of the ERP system), Models are used within specification documents and discussion with MES supplier | Analysis of all IT-systems of the plant, stepwise application on existing production lines | Second MES project for a further production plant |

Additionally, direct modeling during discussions in workshops forces participants to be clear and consistent. Hence, the experts appreciated that an achieved consensus during workshop is directly documented in a comprehensible manner. Further, the experts stated consistently that all modeling elements are required and no element is lacking. One participant suggests an additional iconographic hint for printed information (e.g., barcodes, labels). The general comprehensibility of self-created models was rated with 5.4 points [scale from 1 (very poor) to 6 (very good)]. The comprehensibility of models created from other persons were rated 4.8. Due to its simplicity, models created in one workshop can easily be discussed in other workshops with different participants.

The separation of the three different views (MES/IT, production process and technical system) was assessed very good (5.75 points). The overall usefulness for specification and requirements engineering in MES projects were assessed with 5.4 points.

## VIII. CONCLUSION

This paper presents a novel modeling notation especially for the early phases of MES projects. The MES-ML provides three essential views for a functional definition of MES and its interaction with existing software and automation systems as well as the business and production process. Interactions between different views are formalized by the MES-ML linking model, so that it is possible to integrate the different views and domains in the specification process of MES.

The formal syntax of the MES-ML enables automatic consistency checking of the MES model which increases the model's quality and facilitate modeling.

The benefit of this modeling language is shown taking a yogurt production plant as a case study and is already proved in real MES projects. Experts involved in these industrial evaluation projects estimate that the application of MES-ML leads to a faster specification of MES functionality in correlation to target processes, easier identification of improvement capabilities as requirements for MES implementation and faster identification of ambiguities/inconsistencies of existing processes and data.

Future work includes further application of MES-ML in the industry and first steps towards standardization for comparable specification documents.

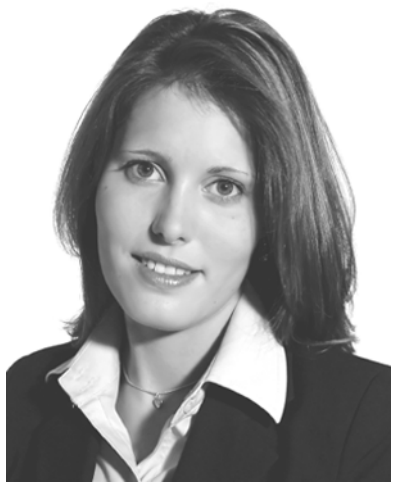

**Maria Witsch** received the Diploma degree in economics (focus on management and production) from Bergische Universität Wuppertal, Germany, in 2006. She is currently working towards the Ph.D. degree at the Institute of Automation and Information Systems, Technische Universität München, Garching bei München, Germany.

Her research interests are in the area of engineering methods with special focus on model-driven (re-)engineering of IT-systems, production systems and business processes as well as vertical and horizontal integration in plant automation.

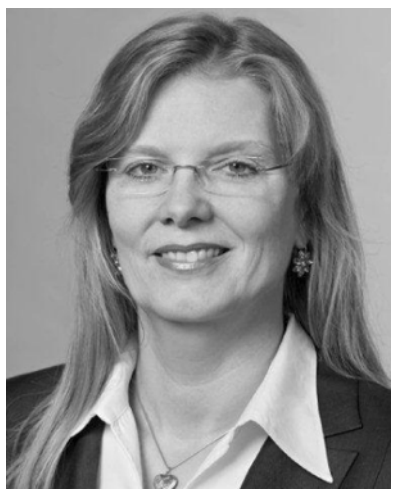

**Birgit Vogel-Heuser** (M'04) graduated in electrical engineering and received the Ph.D. degree in mechanical engineering from the RWTH Aachen, Aachen, Germany, in 1991.

She worked nearly ten years in industrial automation for the machine and plant manufacturing industry. After holding different chairs of automation in Hagen, Wuppertal, and Kassel she is since 2009 Head of the Automation and Information Systems Institute at the Technische Universität München and since 2010 additionally Director for Industrial Automation at FORTISS an affiliated institute of the Technische Universität München. Her research work is focused on modeling in automation engineering for hybrid process and heterogeneous distributed and intelligent systems using a human centered approach.

Prof. Vogel-Heuser is a Member of the GMA (NMO IFAC). She received four awards, i.e., Special Award of the Initiative D21 Women in Research (2005), Borchers Medal of the RWTH Aachen (1991), the GfR Sponsorship Award (1990), and the Adam Opel Award (1989).